\documentstyle[11pt]{article}
\newcommand{\be}{\begin{equation}}
\newcommand{\ee}{\end{equation}}
\newcommand{\beq}{\begin{eqnarray}}
\newcommand{\eeq}{\end{eqnarray}}

\bibliography{unsrt}
\begin{document}
\begin{flushright}
{BROWN-HET-1348} \\
{March 2003}
\end{flushright}
\vskip .20in

\begin{center}
{\bf {TOWARDS A QED-BASED VACUUM ENERGY} }

\vskip .15in
H. M. FRIED\\
\vskip .10in
{{\em Department of Physics \\
Brown University \\
Providence R.I. 02912 USA}}

\vskip .20in
Abstract
\end{center}

A QED-based mechanism, breaking translational invariance of the vacuum at sufficiently small distance scales, is suggested 
as an explanation for the vacuum energy pressure that accelerates the universe.  Very-small-scale virtual vacuum currents are assumed to
generate small-scale electromagnetic fields corresponding to the appearance of a 4-potential $A_{\mu}^{ext} (x)$, which is itself equal to the vev of the
operator $ A_{\mu}(x)$ in the presence of that $A_{\mu}^{ext}(x)$. The latter condition generates a bootstrap-like equation for $A_{\mu}^{ext}(x)$ which has an
approximate, tachyonic-like solution corresponding to propagation outside the light cone, and damping inside; this solution is given in terms
of a mass parameter M that turns out to be on the order of the Planck mass if only the simplest, electron vacuum-bubble is included; if the
muon and tau bubbles are included, M decreases to $\sim 10^6 - 10^7$ GeV.  A multiplicative 4-vector $v_{\mu}$, whose magnitude is determined by a
comparison with the average mass density needed to produce the observed acceleration is introduced, and characterizes the distance d over which
the fields so produced may be expected to be coherent; the present analysis suggests that d can lie anywhere in the range from $10^{-5} cm$
(corresponding to a ``spontaneous vacuum phase change") to $10^{-13}cm$ (representing a ``polarization of the QED vacuum" by quark-antiquark pairs of
the QCD vacuum).  Near the light-cone, such electric fields become large, introducing the possibility of copious charged-particle pair
production, whose back-reaction-fields tend to diminish the vacuum electric field. The possibility of an experimental test of the resulting
plasma at large momentum transfers is discussed.
\vskip .05in
{\bf\section{Introduction}}

	Recent experimental results\cite{one} suggesting the outward acceleration of the universe have concentrated attention on the vacuum as a 
possible source of self-energy, contributing a pressure sufficient to produce the measured acceleration.  In this scenario, Lorentz invariance
(LI) cannot be rigorously true; but since our immediate, measured world is almost perfectly flat, this lack of LI might be expected to play a
minor role.  A major role could be played by the mechanism suggested here, in which a lack of translational invariance (TI) appears when the
virtual, QED vacuum is examined at ultra-short distances.  Those distances range from the inverse of $10^6 - 10^7$ GeV down to the inverse of,
roughly, the Planck mass; the smaller of the two appears if only the simplest, electron vacuum bubble is retained, while the larger is obtained
when the muon and tau vacuum bubbles are included. 

	A constant four-vector $v_{\mu}$ is introduced to characterize the distance over which new, ``effectively external" electromagnetic fields are 
produced and/or may be expected to remain coherent at such very small scales. These fields are generated by electrical currents induced in
the vacuum (in the bootstrap-presence of the very external fields they produce). The order-of-magnitude of $v_{\mu}$ is determined by matching a
derived expression for this vacuum-induced energy density to the mass density $U   \sim 10^{-29} gm/cm^3$ needed to fit observation; \cite{two} in this
computation, the Planck length is used as a natural cut-off, corresponding to the smallest, physically-sensible distances allowed.

	One prejudice of the author should be stated at the outset, which forms the motivation for the present remarks. The by-now conventional 
approach to vacuum energy is that the latter represents, in some fashion, zero-point-energies of relevant quantum fields.  Aside from being
horribly divergent, and so requiring massive renormalizations, it is not clear that zero-point terms even belong in any field theory, for they
can be well-understood as the remnants of improper positioning of products of operators at the same space-time point.  One cannot proceed from
classical to quantum forms without facing this question, which has long ago been given a Lorentz-invariant answer in terms of ``normal
ordering". When Lorentz invariance is subsumed into a more general relativistic invariance, in which all energies couple to the metric, there
is still no requirement of including those zero-point terms which should have been excluded at the very beginning; rather, zero-point terms
have been pressed into service as the simplest way of attempting to understand vacuum energy. Because zero-point energies have been used to
give a qualitative explanation of the Casimir effect and the Lamb shift does not mean that they are the only, or the correct, explanations for
those processes.\cite{three}  It is here suggested that there is another source of vacuum energy, one which violates no principles of quantum field
theory, that can provide a simple and at least qualitatively reasonable QED mechanism for the acceleration of the universe.
\vskip .10in
{\bf\section{Formulation}}

	Our conventional definition of ``vacuum" is the absence of all ``on-shell" phenomena.  Further, there can be no doubt of the reality of ``off-shell" vacuum fluctuations, demonstrated long ago by the -27 megacycles/sec contribution to the $^{2}S_{1/2} - ^{2}P_{1/2}$ Lamb shift of Hydrogen.\cite{four}  As formulated by well-known axioms of QFT, the vacuum state contains no on-shell particles; and for QED, in particular, the vacuum expectation value $(vev)$ of the current operator $j_{\mu} µ(x) = ie \vec{\psi} (x) \gamma_{\mu} \psi (x)$
                    (renormalization and split-point arguments aside) must vanish in the absence of classical, external fields, designated by $A_{\mu}^{ext}(x)$,
\be
\langle 0 \vert j_{\mu} (x) \vert 0 \rangle_{A^{ext} = 0} \, = \, <\langle 0 \vert j_{\mu} (0) \vert 0 \rangle_{A^{ext} = 0 } \, = \, 0\, .
\ee
										
When such a classical $A_{\mu}^{ext}$ is present, the current it induces in the vacuum can be non-zero,		
$$
\langle 0 \vert j_{\mu} (x) \vert 0 \rangle_{A^{ext} \not=} \not\equiv 0 ,
$$
although strict current conservation demands that\\
 $\partial_{\mu} <0\vert j_{\mu} (x) \vert 0 > _{A^{ext}\not= 0} = 0$. 

	The conventional mathematical apparatus used to describe the vacuum state, or vevs of current operators, 
makes no reference to the scales on which vacuum properties are to be observed; and in this sense, it is here suggested that the conventional
description is incomplete.  On distance scales larger than the electron Compton wavelength, $\lambda_e \sim 10^{-10} cm$, one can imagine that the average
separation of virtual $e+$ and $e-$ currents is not distinguishable, and hence that the vacuum displays not only a zero net charge, but also a zero
charge density.  But on much smaller scales, such as $10^{-20} cm$, the average separation distances between virtual $e+$ and $e-$ are relatively
large, with such currents describable as moving independently of each other - until they annihilate.  Since there is nothing virtual, or
``off-shell", about charge, on sufficiently small space-time scales such "separated"  currents can be imagined to produce effective "external" 
fields, characterized by an $A_{\mu}^{ext}(x)$, which could not be expected to be measured at distances larger than $\lambda_e$, but which exist - and
contain electromagnetic energy - on scales much smaller than  $\lambda_e$.

	We therefore postulate that, at such small scales and in the absence of conventional, large-scale $A_{\mu}^{ext},<0\vert j_{\mu} (x) \vert 0>\,$ need be neither $x$-independent nor zero; but, rather, that it generates an $A_{\mu}^{ext}(x) \,$ discernible at such small scales, which is
given by the conventional expression

\be
A_{\mu}^{ext} (x) = \int d^4 y D_{c,\mu\nu} (x-y) <j_{\nu}(y) >\, ,
\ee							
where$ D_{c\mu\nu}$  is the conventional, free-field, causal photon propagator that, for convenience, is defined in the Lorentz gauge,
$$
0 = \partial_{\mu} A_{\mu}^{ext} = \partial_{\mu} D_{c,\mu\nu} = \partial_{\nu} D_{c,\mu\nu} \, .\nonumber
$$

	For comparison, note that classical electromagnetic vector potentials can always be written in terms of well-defined, classical 
currents $J_{\mu}$        , by an analogous relation, ${\cal A}_{\mu} = \int D_{c,\mu\nu} J_{\nu} ,$  
while the transition to QED in the absence of conventional, large-scale, external fields involves the replacement of 
the classical vector potential and currents by operators $A_{\mu}(x)$ and $j_{\mu}(x) = ie\bar{\psi} (x) \gamma_{\mu} \psi (x) ,$  
which satisfy operator equations of motion, $(-\partial^2 ) A_{\mu} = j_{\mu},$  
or															    \be
A_{\mu} = \int D_{c,\mu\nu} <j_{\nu}> + A_{\mu}^{IN} \, .
\ee
Calculating the vev of (2.3) yields

\be
< A_{\mu} > = \int D_{c,\mu\nu} < j_{\nu} > \, ,
\ee
and conventionally, in the absence of the usual, large-scale external fields, both sides of (2.4) are to vanish.  
However, if we assume that non-zero \\
 $<0\vert j_{\mu} (x) \vert 0>$  can exist on ultra-short scales, then a comparison of (2.4) with (2.2) suggests that the $A_{\mu}^{ext} (x)$
produced by such small-scale currents are to be identified with $<0\vert A_{\mu} (x) \vert 0> $ found in conventional QED in the presence of the same $A_{\mu}^{ext} (x)$.  In
other words, 
\beq
A_{\mu}^{ext} (x) = < A_{\mu} (x) >& = & {1\over i}  \int  D_{c,\mu\nu} (x-y) {\delta\over\delta A_{\nu} (y)} \cdot e^{-{i\over 2} \int {\delta\over\delta A} D_{c} {\delta\over\delta A}}\cdot \nonumber\\
& & {e^{L[A+A^{ext}]}\over < S[A^{ext} ]>} \vert_{A\rightarrow 0} ,
\eeq
which provides a bootstrap equation with which to determine such short-scale $A_{\mu}^{ext} (x)$, 
if any exist.  In (2.5), which can be transformed into a functional-integral relation, the vacuum-to-vacuum amplitude is given\cite{five} by 
\be
< S [A^{ext} ] > = e^{-{i\over 2} \int {\delta\over\delta A} D_{c} {\delta\over\delta A} } \cdot e^{ L[A+A^{ext} ]} \vert_{A\rightarrow 0} ,
\ee
with $L[A] = Tr \ln [1-ie \gamma \cdot   A S_c  ]$. 

	Of course, one immediate solution to (2.5) is $A_{\mu}^{ext}(x) = 0 = < A_{\mu}>$, 
the conventional solution.  But we are interested in, and shall find, solutions that may be safely neglected at conventional nuclear and
atomic distances, but which are non-zero in an interesting way at much smaller distances.  In a sense, such non-zero solutions are akin to
those found in symmetry-breaking processes, such as spontaneous or induced magnetization; qualitatively similar ideas have previously been
discussed elsewhere,\cite{six} for other reasons. We argue below that these short-scale $A_{\mu}^{ext}(x)$ could be spontaneous, or induced by other, virtual,
QCD processes.

\vskip .10in

{\bf \section{Approximation}}

	How does one go about finding a solution to (2.5)?  The first requirement is a representation 
for $L[A+A_{\mu}^{ext}]$ which is sufficiently transparent to allow the functional operation of (2.5) to be performed; but, sadly, this is still a distant goal, which has been approximately realized in only a few, special cases.\cite{seven}  
What shall first be done here is to use the simplest (perturbative) approximation to $L$, 
$$
L[B] \Rightarrow {i\over 2} \int B_{\mu} K_{\mu\nu}^{(2)} B_{\nu} \, ,
$$
where the renormalized\cite{four} $K^{(2)}_{\mu\nu}$     corresponds to using only the simplest, order-$e^2$, closed-electron-loop,
$$
\tilde{K}_{\mu\nu}^{(2)} (k) = \left( \delta_{\mu\nu} - k_{\mu}k_{\nu}/k^{2} \right) \cdot k^2 \pi (k^2 ) , \,\, k^2 = \vec{k}^2 - k_0^2 \, ,\nonumber
$$
and
\be
\pi (k^2) = {2\alpha\over\pi} \int_0^1 dx \cdot x (1-x) \ln \left( 1 + x (1-x) {k^2\over m^2} \right) \equiv {2\alpha\over\pi} \phi \left( k^2/m^2 \right) \, ,
\ee
where      $\alpha  = e^2/4\pi \simeq    1/137$ denotes the renormalized coupling, and $m$ is the electron mass. 
 Higher-order perturbative terms should each yield a less-important contribution to the final answer, although the latter could be
qualitatively changed by their sum; we shall assume that this is not the case, and that the perturbative approximation (properly unitarized by
the functional calculation which automatically sums over all such loops) gives a
qualitatively reasonable approximation.

The functional operation of (2.5), now equivalent to Gaussian functional integration, is immediate and yields the approximate relation for this $A_{\mu}^{ext} (x)$,
$$
A_{\mu}^{ext} (x) = \int d^4 y \left( D_c K {1\over 1-D_c K} \right)_{\mu\nu} (x-y) A_{\nu}^{ext} (y) \, ,
$$
or
\be
\tilde{A}_{\mu}^{ext} (k) = \left( \pi (k^2 ) {1\over 1-\pi (k^2 )} \right) \, \tilde{A}_{\mu}^{ext} (k) \, .
\ee										
A non-zero solution to (8) may be found in the ``tachyonic" form
\be
\tilde{A}_{\mu}^{ext} (k) = C_{\mu} (k) \delta (k^2 - M^2 )
\ee
which then requires that
$$
1 = \pi (M^2 ) \, {1\over 1 - \pi (M^2 )} \, , \, \, \pi (M^2 ) = 1/2 \, ,
\nonumber
$$
or $\phi ({M^2\over m^2}) = {\pi\over 4\alpha} $, which serves to determine $M$.     Note that a solution 
of form $C_{\mu} \delta (k^2 + \mu^2 )$  would not be possible, since the log of $ \pi (-\mu^2 )$ picks up an imaginary contribution for time-like $k^2 = - \mu^2 $ , for $\mu > 2m$.  

	An elementary evaluation of the integral of (7) for large $M/m$ yields
\be
\phi \left( {M^2\over m^2 } \right) \simeq \ln \left( 1 + \left( {M\over 2m} \right) ^2\right) + 0 (1) \sim 2 \ln \left( {M\over 2m} \right) + \cdots ,
\ee
so that $M \simeq  2m \,exp[\pi /8 \alpha ]$, neglecting relatively small corrections.  
Since $\pi /8 \alpha \simeq 53.8$,  (3.10) yields a value for $M$ close to the Planck mass, 
$M \sim 2\times 10^{20} GeV/c^2$.  Because the Physics at distances $\sim M_P^{-1}$ - and, indeed, 
the continuum nature of space and time - is still a matter of speculation, where processes well-defined at larger scales are no longer 
meaningful, one must exclude intervals smaller than $M_P^{-1}$ when calculating any quantity, such as the electromagnetic energy contained in the
fields corresponding to the $A_{\mu}^{ext}(x)$ of (9).

	If the muon and tau self-energy bubbles are included,\cite{eight} 
the  numerical situation changes, and somewhat more favorably, because the value of $ M $ found above decreases to between $10^6$ and $10^7\, GeV$;
this is still large enough to correspond to wave-numbers much larger than those associated with atomic or nuclear distances. If $M_0$ denotes the
previous value of this parameter calculated as above, using only the electron bubble, when one includes muon and tau bubbles there results a
new, approximate formula for $M$, given by 
$$			
M = (m_{\mu} m_{\tau}M_0)^{1/3}\,.   
$$
However, the upper cut-off on the integral of (14), below, is still the Planck mass, $M_P$,
 and generates the same order-of-magnitude from which the possible $d$-values, below, are deduced.  

	The reason that quark-antiquark vacuum-polarization bubbles have not been included is simply 
the author's prejudice that those processes would be dominated by strong-coupling gluonic effects, which always operate on time-scales much
shorter than those of QED; and hence one might expect that $q-\bar{q}$ vacuum bubbles would be irrelevant to this computation. 

	It is also quite difficult to estimate such effects, taking into account the appropriate gluonic structure.  
Of course, one might argue that at very short distances, the gluonic effects are suppressed by asymptotic freedom, and hence one could employ a
$q-\bar{q}$ bubble resembling the leptons', except for a change of charge.  But what does one then use for the quark mass?  Asymptotic states are not
specified in terms of the quark mass, and a calculation that employs $m_q$ should be a simplifying approximation to much more complicated
Physics.  Were $m_q$ here chosen as one-third of the nucleon mass, then there will be some diminution in the overall $M$, but nothing of
consequence.  If $m_q$ is taken as very small, there is then the danger that $\pi (M^2) $          can be larger than unity, and there would be no non-zero
solution to (8)  - which might be the desired conclusion of those who would prefer that such an $A_{\mu}^{ext} (x)$  be identically zero, and that the
remaining analysis can be dispensed with! 

	A specific choice of an effective $m_q$, however, really corresponds to a statement concerning 
the space-time region over which one expects an analysis to be valid; and from this point of view, a choice of $m_q$ less than the electron's mass
- which can lead to an overly large value of  $\pi (M^2 )$          - makes no physical sense because it suggests that QCD effects can appear at distances
larger than $10^{-13} cm$.  Until one reaches the cut-off distances signified by $M_P^{-1}$, it would seem most appropriate to include the full gluonic
structure of QCD in any  matrix element; and here one has the observed fact that strong-interaction processes are much more rapid than those of
QED, which in the present case one might interpret as the idea that such $q-\bar{q}$ bubbles, containing gluonic forces between $q$ and $\bar{q}$, appear and
disappear so rapidly that their effect on the present QED calculation can be ignored (except for the possibility, mentioned below, that such
rapid QCD fluctuations can serve to provide the initial impetus for the QED bootstrap process). This, in addition to one's ignorance of the
proper way of characterizing $m_q$ in quark bubbles without gluonic binding, is the authors's justification for the neglect of such ``simple" $q-\bar{q}$
vacuum bubbles.
\vskip .10in

{\bf\section{Computation}}

	In order to fully describe this $A_{\mu}^{ext}$ , one must define $C_{\mu}(k)$.   
The simplest choice would be $ C_{\mu} \sim   k_{\mu}$; but this corresponds to a ``pure gauge", leading to zero $F_{\mu\nu}$.  If $ C_{\mu} \not= 0$, it must depend on some other
4-vector $\omega_{\mu}$   , either of origin external to this problem, or of origin within QED, corresponding to an effective phase change.  This question
will be left open until the determination of the magnitude of $C_{\mu}$ is made, in comparison with the average value of mass density needed to
account for the outward acceleration of the universe.

	For the computation of $A_{\mu}^{ext}(x)$, we shall simply choose $C_{\mu} (k) = v_{\mu}$, where $v_{\mu}$ is a constant 4-vector (of unknown origin), and shall try 
to justify that choice below.  Note that $C_{\mu} \rightarrow k_{\mu} - v_{\mu} (k\cdot v ) /k^2 $ 		   would be the more appropriate choice in order to retain the Lorentz gauge condition; but
since the additional term is ``pure gauge", it does not contribute to the fields, and will be dropped.  In the familiar and convenient
``particle" units of $c = \hbar =1, v_{\mu}$  has the dimensions of length, or inverse mass. It will subsequently be useful to state the dimensions of $v_{\mu}$ 
in a more general way; and a simple, dimensional inspection shows that the correct dimensions of $C_{\mu}$ are (length)$^{3/2}$ multiplied by (energy)$^{1/2}$, 
so that $v_{\mu}$ may be written as 
\be
\epsilon_{\mu} \, d^{3/2} \left( \mu c^2 \right)^{1/2} = \epsilon_{\mu} \, d^{3/2} \left( {\mu\over m} \right)^{1/2} (mc^2)^{1/2} \, ,
\ee										
where $\epsilon_{\mu}$      is a 4-vector of unit magnitude, $\epsilon^2 = \pm     1$.  If $c=1$, and $ \mu = m$,  
then an average value of d (calculated from an average value of the magnitude of $v$) is    $\sim 10^{-5} cm$; but if $\mu = M_P$, then the same, average
magnitude of $v$ means that $d$ is much smaller.  The magnitude of $v$ will be determined by matching the average energy density of the
electromagnetic fields obtained from $A_{\mu}^{ext}(x)$ to the experimental $U $ determined from astrophysical data.\cite{two}  These two extreme values of $d $ may
be thought of as defining the scale at which the vacuum fields are generated, or the spatial size at which they can be maintained.  If these
fields of very high wave-number are spontaneously created, $d$ can turn out to be on the order of $10^{-4} cm$, while if these fields are induced by
virtual QCD processes, $d$ can be on the order of $10^{-13} cm$.  The present calculation cannot distinguish between these possibilities, since any
solution of (8) that is multiplied by an arbitrary constant is still a solution.

Of course, the simple replacement $C_{\mu} \rightarrow v_{\mu}$   is insufficient to 
provide the physical cut off at frequencies $\sim M_P$; but this could easily be remedied by adjoining a factor $\theta (M_P - \vert k_0 \vert )$     		   to  $v_{\mu}$ in (3.9).  Also, as
a matter of principle, when considering an expanding universe of age $T$, it makes no (conventional) quantum mechanical sense to include
frequencies, or energies, ``softer" than $T^{-1}$; were such a cut-off needed, it could be achieved by also inserting a factor of $\theta (\vert k_0 \vert - T^{-1} )$ in (9).  What
shall be done instead is to follow that simplest procedure which uses $C_{\mu} = v_{\mu}$ in order to obtain a qualitative understanding of $A_{\mu}^{ext}(x)$ , and
its corresponding $ E $ and $B$ fields; but, then, when the total energy of those fields is calculated, to introduce the corresponding cut-offs, $M_P$ 
and $T^{-1}$, in the integral (over all space at a given time) of the total electromagnetic energy,	

	Integration of the Fourier transform of (3.9), with $ C_{\mu}\rightarrow v_{\mu} $ , is straightforward, 
and yields the ``tachyonic" forms,
\beq
A_{\mu}^{ext} (x) & = & - {2\pi^2 M\over {\sqrt x^{2}}} \, v_{\mu} N_1 \left( M\sqrt{x^2} \right) , \, x^2 > 0\nonumber\\
& = & - {4\pi M\over\sqrt{-x^2}} \, v_{\mu} \, K_1 \left( M \sqrt{-x^2 }\right) , \, x^2 < 0 \\
& = & 0 , \,\, x^2 = 0 \, , \nonumber
\eeq
where $x^2 = \vec{x}^2 - x_0^2 $ , with regions of propagation and damping, inside and outside the 
light cone (l.c.), reversed in comparison to those of a causal problem.  [The third entry of (4.12) represents the Principal Value limit
of $1/x^2$  , as  $x^2\rightarrow 0$.]   As the l.c. is approached, from inside or outside, $A_{\mu}(x)$ becomes arbitrarily large at intervals $\sim M_P^{-1}$ 
from the l.c.; but, as previously noted, quantum gravity - or what may be the same thing, a lack of continuous space-time variables - sets a
practical limit to all processes at such high frequencies.  Incidentally, outside the l.c. one makes a distinction between frequencies, $k_0$, 
and wave-numbers, $ \vec{k}$, because $\delta (k^2 - M^2 ) = \delta (\vec{k}^2  - [ k_0^2 + M^2 ] )$  , so that $\vert \vec{k}\vert$ is at least as large as $M$ , while $k_0$ runs from $T^{-1}$ to $M_P$

	Inspection of (2.2), with a conserved $<0\vert j_{\mu} (x) \vert 0 >$,  suggests that one can instead write
$$
A_{\mu}^{ext} (x) = \int D_c (x-y) < J_{\mu} (y) > \, ,\nonumber
$$
and for the solutions of (11), it then follows that
$$
\left( - \partial^2 \right) A_{\mu}^{ext} (x) = M^2 A_{\mu}^{ext} (x) = < j_{\mu} (x) > \,. \nonumber
$$

Hence this $<j_{\mu}(x) >$, as a function of any coordinate system used to describe it, 
has propagation only outside the l.c. of that system. A most unusual current!  
Real currents of real particles propagate inside l.c.s.

	When the $A_{\mu}^{ext}(x)$ of (4.12) is used to calculate the corresponding $E$ and $B$ fields, and the latter are inserted into $ U = {1\over 8\pi} ( E^2 + B^2 )$  , one finds unavoidable singularities as the l.c. is approached; these singularities are
effectively removed by refusing to consider space-time intervals $< M_P^{-1}$.  The computation is simpler in momentum space, if we calculate an
average energy density defined as the total vacÙum electromagnetic energy divided by the total volume of our universe,
$$
\bar{U} = {1\over 8\pi} \int d^3 x \left( E^2 + B^2 \right) / {4\pi\over 3} R^3 \, , \nonumber
$$
and if we simplify matters by integrating $\int d^3 x$ over all space, rather than fixed by $\int d^3 x = 4 \pi R^3/3$.   
One begins with the Fourier representation of (3.9), with  $C_{\mu} \rightarrow  v_{\mu}$ ; and one then calculates 
$$
W = \int d^3 x U\nonumber
$$
over all space at a fixed time.  This computation requires the specification of      $v_0^2$ and the  $v_i^2    $ , 
which enter on the same, multiplicative footing; and to make the calculation even simpler, we imagine that       $ B = 0, E \not= 0$, so that only
electric fields are present.  One then finds
\be
W\Rightarrow 4 \pi^3 v_0^2 \int_0^{\infty} \, {dk_0\over k_o} \left( k_0^2 + M^2 \right)^{3/2} \cos ^2 (k_0 x_0 )\,
\ee											and for  $x_0 >>M^{-1}$ , replace the $\cos^2 (k_0 x_0 )$ factor by 1/2, and insert upper and lower limits to the integral of (13), replacing it by
\be
W \rightarrow 2 \pi^3 v_0^2 \int_{T^{-1}}^{M_P} \, {dk_0\over k_0} \left( k_0^2 + M^2 \right)^{3/2} \, .
\ee												
From (14) it is clear that the dominant effect in any Lorentz frame comes from the region outside of, 
but on the order of a distance $1/M_P$ from the l.c.; and this generates a result for $W$ proportional to $(M_P)^3$. It is amusing to note that this is
one power of cut-off less than would be needed in a computation which sums over all permitted, zero-point energies - but, of course, the
Physics here is quite different.  One finds
$$
W \sim {2\over 3} \pi^3 v_0^2 M_P^3 + \cdots\nonumber
$$
which when divided by 
$$
{4\pi\over 3} R^3 \sim 4 \times 10^{85} cm^3
$$
produces a crude, average value for the total energy density (or mass density, when divided by $c^2$).  
When this ratio is equated to the average mass density needed to produce the observed positive acceleration of the universe$,  U \sim    10^{-29} 
gm/cm^3$,  one finds the value of $\sim 10^8$ for the dimensionless quantity   $v_0m$.  

	From this numerical value follow the values quoted for $d$ in the discussion following (4.11).  From the relation 
$$
v_0 \sim d^{3/2} \mu^{1/2} \, ,\nonumber
$$
the range of $\mu$ values, from $m$ to $M$ to $M_P$, define different, possible values of $d$, 
which may be interpreted as the length over which the electric fields are coherent, with a minimum uncertainty of time duration on the order of
magnitude of $t = 1/\mu$; in these estimates, we assume that, once an $A_{\mu}^{ext}$ appears, it is to last indefinitely (except for the diminution of 
fields discussed in Section V.).  Table 1 relates these approximate, average quantities.
\begin{center}
\vskip .10in
\begin{tabular}{ |  l | l |}\hline
 $\mu $ &  $d(cm)$  \\ \hline
 $m $ &  $10^{-5}$  \\  \hline
$M $ &  $10^{-8}$  \\ \hline
 $M_P$ & $10^{-13}$ \\ \hline
\end{tabular}
\vskip .10in

Table 1:  Approximate orders of magnitude of $d$ 
for different choices of $\mu$
\end{center}
\vskip .10in

	If $\mu \sim m$, then the picture that appears is one of weak fields extending over distances on the order of $10^{-5} cm$; 
and one might think of these fields as occurring spontaneously, in the sense that there is no outside influence that initially drives them.  At
the other extreme, if $\mu\sim M_P$, then $d \sim 10^{-13} cm$, a typical, strong-interaction distance.  It is then tempting to interpret  $d$ as the scale at
which such $A_{\mu}^{ext}(x)$ are generated, by a physical, albeit virtual, process that is external to QED.  Here, the obvious candidate is QCD; and one
can imagine, as part of the QCD vacuum structure, virtual quark-antiquark pairs, electrically charged, forming and reforming, with a
partially-formed gluonic flux tube between them. This QCD-generated separation of charge, on the order of $10^{-13} cm$, effectively ``polarizes the
QED vacuum", and in turn produces a non-zero $A_{\mu}^{ext}$, which the rules of QED require to satisfy (3.8). 

	Other models, lying in between these two extreme cases are surely possible, 
but can only be determined by a more-accurate calculation, one which invokes QED vacuum structure more complicated than the simplest,
closed-fermion-loop of this paper.  But, then, the initial, functional part of the calculation can only be carried through with yet another,
unjustified, approximation.  In a better calculation, $v_{\mu}\rightarrow v_{\mu} (x) $, and would be determined by the passage from the exact (2.5) to an improved
(3.8), involving higher powers of $A^{ext}$ on the RHS of the improved (3.8).

	The question has also been raised [8] as to how much significance can 
be attached to the replacement of $C_{\mu}(k)$ by a constant $v_{\mu}$; and the
response here is somewhat more involved, and ties in with the form of the calculation performed. The choice of constant $v_{\mu}$ was originally made
for simplicity, for it permits a simple evaluation of all relevant integrals.  But if $d \sim   10^{-5} cm $ corresponds to spontaneously-appearing
fields in a volume of that dimension cubed, then there should exist other ``domains" with other values of $v$. However, if the direction of the
spatial $v$-vector is random - and one would expect this if the process is spontaneous - then an average over the $v$-directions can produce the
same, final effect, since the energies of each region will just add.  Since there is no reason to expect different regions of space-time (that
are matter-free) to be different, one would also expect that the magnitudes of $v_0$ and $\vec{v}$ will be the same.; and this ``justifies" the constant-$v$
calculation.

	If, however, $d  \sim  10^{-13} cm$, and the entire effect is driven by QCD 
polarization of the QED vacuum, then the picture becomes somewhat more reasonable.  Now the appearance of a ``constant" $v_{\mu}$ providing the
4-vector index for $A_{\mu}$ no longer raises the question of LI violation,  since $v_{\mu}$ is itself ``dynamical", defined by the average separation of
virtual $q-\bar{q}$ pairs; rather, the question now becomes one of a proper characterization of the QCD vacuum. What emerges from this ``tachyonic"
solution is the intruiging idea that vacuum dynamics is somehow connected with events that are outside the l.c., in contrast to our
conventional, causal world, whose dynamics are always on or inside the l.c.. It should be emphasized, however, that even if conventional LI is
rigorously maintained when details of the QCD vacuum become known, conventional translational invariance of the vacuum is open to question (and
redefinition).

	Another question has been raised\cite{nine} concerning the choice of origin of the coordinate system used.  
If observer $A$ uses a coordinate system with spatial origin at the tip of the Eiffel Tower, while observer $B$ uses a coordinate system with
spatial origin fixed at the center of Brown University, then one expects that any $U$ calculated at an intermediate point, at the same time, will
display different values to $A $ and $B$, because their separation distance $D     \simeq 5000 km$.  What one might do is to replace $x$ by $x-x_0$, treat $ v
=v(x_0)$, and average over all $x_0$.  But such a step is not necessary for the present, crude estimates, because the major contributions to $U$ (at a
given time) come from distances close to $1/M_P$ from the l.c.; and even if this were not the case, when $U$ is averaged over the entire universe of
spatial radius $R >>>> D$, both $A$ and $B$ should obtain the same (averaged) answer.
\vskip .10in

{\bf\section{Detection}}

	Let us assume that the electric fields following from the $A_{\mu}^{ext}$ of (3.9) 
are fairly decent representations of what occurs physically, and consider the situation near the l.c., when $E^2 >> m^4/\alpha$    , and significant pair
production can take place.  The most relevant formula is then that of Schwinger's 1951 paper\cite{five} for the vacuum persistence probability $P_0$,
\be
\ln P_0 = -{Vt\alpha E^2\over \pi^2} \, \sum_{n=1}^{\infty} \, {1\over n^2} \, e^{-n\pi m^{2}/eE}
\ee
where $\vec{E}$ is a constant electric field across a volume $V$ which has existed for a time $t$.  In the limit of very large $\vert\vec{E}\vert = E$, this sum can be trivially evaluated as 
$$
\ln P_0 = -\Gamma \cdot t \, , \, \Gamma = {VeE m^2\over 4 \pi^2 } \cdot \kappa \, , \, \kappa = \int_1^{\infty} \, {dx\over x^2} e^{-x} \sim 0 (1) \, ,
\nonumber
$$ 
so that the vacuum state would disappear exponentially fast into a state of many lepton pairs.  

	What such a computation misses is the effect of the back-reaction, or oppositely-induced 
field generated by the emitted pairs, which tends to reduce the original $E$.\cite{ten} In the present case, and even though the fields are not
constant, near the l.c. one would expect copious lepton production, leading to a sort of averaged decrease of $E$, and an effective plasma, or
``thermalization" of the combination of effective $E$ and number $n$ of lepton pairs.  (The previous computation of electrical energy density
``stored" in the vacuum is essentially unchanged, except that that energy is now partitioned between weaker fields and the lepton plasma.) 
However, as the l.c. is approached, both the effective $ E$¸ and the number $n$ of pairs must tend to zero, because of the ``experimental" proof
that a photon on the l.c. displays no interaction with either an effective $E $ nor with the lepton plasma. 

	Could the existence of such a plasma near the l.c. be tested?  
Consider the simple example of charged-particle scattering by the exchange of a single, virtual photon.  The amplitude for this process is
proportional to $(q^2 - i\epsilon  )^{-1}$, and in the center of mass of the scattering particles, $q_0 = 0$.  Our solution for $A_{\mu}^{ext}$ becomes large near the
l.c. when $M^2 x^2 < 1$, and its Fourier transform is proportional to $ \delta (q^2 - M^2 ) = \delta (\vec{q}^2 - M^2 )$	for $q_0 = 0$.  In other words, $\tilde{E}(q)$ is peaked when $\vert\vec{q}\vert    = M$,  
which is where the plasma region begins, and where absorption or rescattering by the plasma of this or any other virtual photon would occur.  
Hence the presence of such a plasma could be tested - if one could ever reach momentum transfers on the order of $M$ - by observing an anomalous
drop in the differential cross section when $\vert \vec{q}\vert$ exceeds $ M$.

This observation raises the question of the effect of such vacuum structure on the typical $UV$ divergences of QED.  If all virtual quanta at large momenta can be partially absorbed or rescattered by these vacuum fields, or by the plasma resulting from their essentially non-perturbative pair-production, one can expect an effective damping, or built-in convergence factor, for all sufficiently large momenta $ M< q< M_P$.  In other words, a redefined vacuum could provide the necessary convergence factors to insure finite QFT perturbation expansions.  The essential ideas are that such vacuum fields exist in any and every Lorentz frame; and that they become significantly large as the l.c. is approached.

\vskip .10in
{\bf\section{Evaluation}}

	There are other, less-dramatic and numerically less-important solutions to (3.8), 
which might be mentioned.  For example, the choice
$$
C_{\mu} (k) = \xi v_{\mu} \delta (k \cdot v)\nonumber
$$
rigorously maintains the Lorentz gauge condition; and, with $v_{\mu}$ timelike, 
has no singularity on the l.c., and is non-zero only outside the l.c., 
$$
A_{\mu} (x) = \xi {v_{\mu}\over\sqrt{-v^2}} \, M \phi (MX) , \phi (u) = {\sin u\over u} \, ,\nonumber
$$
where $X = (x^2 - (x\cdot v)^2/v^2)^{1/2}$ and $\xi$    is a constant.  
Rotating $v_{\mu}$ to lie along the 4-direction produces the simple, time-independent result, $A_{\mu}(x) \rightarrow i A_0 = \xi (2\pi/r)\sin (Mr)$     ,  with $r = \vert\vec{x} \vert$.  In order to have fields from such a solution generate the needed acceleration of the universe,  $\xi$     would have to be chosen extremely
large.

	Whichever the form of appropriate $ A_{\mu}^{ext}$ chosen, 
the essential and new physical idea proposed above is that on sufficiently small space-time scales, induced vacuum currents should generate 
effective, external fields, fields that are associated with vacuum structure, and have measurable consequences.  It is difficult not to 
imagine that these electromagnetic fields, residing principally outside any l.c., are the source of the astrophsicists' missing, ``dark
energy", while the lepton plasma they must surely produce in regions near the l.c. represents the missing ``dark matter".   

	One cannot deny that the calculations and estimates above are crude. 
Nevertheless, this mechanism for generating a significant vacuum energy from  electromagnetic fields of very high wave numbers and
frequencies may have a certain reality.  One should realize that the energy produced by such virtual currents is real; and that crudeness does
not necessarily preclude correctness.


\begin{thebibliography}{**}
\bibitem{two} The author is indebted to I. Dell¹Antonio for this numerical estimate, as well as for several illuminating conversations.
\bibitem{nine}  The author is indebted to H. de Vega for this, as well as other, 
most-constructive criticism.
\end{thebibliography}
\end{document}